\documentclass[prl,twocolumn,amsmath,amssymb,
superscriptaddress,showpacs,floatfix,nobibnotes,noeprint]{revtex4-1}
\usepackage{graphicx}
\usepackage{amsmath}
\usepackage{soul,xcolor}
\usepackage{color}
\usepackage{placeins}
\usepackage{braket}
\usepackage[normalem]{ulem}
\usepackage{verbatim}
\usepackage{natbib}

\usepackage{mathtools}

\definecolor{dgreen}{rgb}{0.0, 0.5, 0.0}

\begin{document}

\preprint{APS/123-QED}

\title{Probing Nonlocal Spatial Correlations in Quantum Gases with Ultra-long-range Rydberg Molecules}
%Mean-Field, Classical, and Quantum Many-Body Descriptions of  Rydberg Polarons  in a Bose Gas}
%\title{First-Principle Derivation of Mean-Field, Classical, and Quantum Many-Body Descriptions of  Rydberg Polarons  in a Bose Gas}% Force line breaks with \\
%%\thanks{A footnote to the article title}%
%\author{The Usual Suspects}

\author{J.\,D. Whalen}%joseph.d.whalen@rice.edu
\author{S.\,K.\,Kanungo} %soumya.k.kanungo@rice.edu
\author{R. Ding}% Roger.Ding@rice.edu
\affiliation{
 Department of Physics \& Astronomy and Rice Center for Quantum Materials, Rice University, Houston, TX 77251, USA
}

\author{M. Wagner}% marcel.wagner@mpq.mpg.de
\author{R. Schmidt }
% richard.schmidt@mpq.mpg.de

\affiliation{Max-Planck-Institute of Quantum Optics,
Hans-Kopfermann-Stra{\ss}e 1, 85748 Garching, Germany}
\affiliation{Munich Center for Quantum Science and Technology,
Schellingstra{\ss}e 4, D-80799 M\"{u}nchen, Germany
}
%\author{E. Demler }
%%% demler@physics.harvard.edu \homepage{http://www.Second.institution.edu/~Charlie.Author}
%\affiliation{
%Department of Physics, Harvard University, Cambridge, MA 02138, USA
%}%

\author{H.\,R. Sadeghpour }
%% hrs@cfa.harvard.edu \homepage{http://www.Second.institution.edu/~Charlie.Author}
\affiliation{
ITAMP, Harvard-Smithsonian Center for Astrophysics, Cambridge, MA 02138, USA
}%

\author{S. Yoshida} %shuhei@concord.itp.tuwien.ac.at
\author{J. Burgd\"orfer}%burg@concord.itp.tuwien.ac.at
\affiliation{%
Institute for Theoretical Physics, Vienna University of Technology, Vienna, Austria, EU
}%

%%\collaboration{MUSO Collaboration}%\noaffiliation
\author{F.\,B. Dunning}%fbd@rice.edu
\affiliation{
 Department of Physics \& Astronomy and Rice Center for Quantum Materials, Rice University, Houston, TX 77251, USA
}

\author{T.\,C. Killian}%
\email{corresponding author: killian@rice.edu}
\affiliation{
 Department of Physics \& Astronomy and Rice Center for Quantum Materials, Rice University, Houston, TX 77251, USA
}
\date{\today}% It is always \today, today,

\begin{abstract}
{We present photo-excitation of ultra-long-range Rydberg molecules  as a probe of spatial correlations in quantum gases. Rydberg molecules can be created with
well-defined internuclear spacing, set by the radius of the outer lobe of the Rydberg electron
wavefunction $R_n$. By varying the principal quantum number $n$ of the target Rydberg state, the molecular excitation rate can be used to map the pair-correlation function of the trapped gas $g^{(2)}(R_n)$. We demonstrate this with ultracold  Sr gases and probe pair-separation length scales ranging from $R_n=1400-3200$\,$a_0$, which are on the order of the thermal de Broglie wavelength for temperatures around 1\,$\mu$K.
%By varying the principal quantum number of the target Rydberg state from $n=30-45$ in ultracold strontium gases, the excitation rate is used to map the pair-correlation function  for lengthscales from $1000-3000$\,$a_0$, which is on
We observe bunching for a single-component Bose gas of $^{84}$Sr and anti-bunching due to Pauli exclusion at short distances for a polarized Fermi gas of $^{87}$Sr, revealing the effects of quantum statistics.
}
\end{abstract}

\maketitle

%\tableofcontents

%\section{\label{Sec:Intro}Introduction}

%\bibliography{bibliography}% Produces the bibliography via BibTeX.
%\bibliography{C:/Users/killian/Documents/PAPERS/BIBLIOGRAPHY/bibliography,C:/Users/killian/Documents/PAPERS/BIBLIOGRAPHY/MENDELAYTEMPLIBRARYBIBFILE/library,bib_RydMol}
Our understanding of quantum gases has been greatly enhanced by \textit{in situ} measurements of spatial correlations, which can arise from  Bose or Fermi quantum statistics \cite{ngl99,bgm97,kmi97,zhg03,obh15} or the formation of more complex entangled states \cite{bdz08,loh04,kww04}.
%\textit{In situ} measurements of spatial correlations have greatly enhanced the understanding of quantum gases because correlations, such as those arising from  Bose or Fermi quantum statistics \cite{ngl99,bgm97,kmi97,zhg03,obh15} or the formation of more complex entangled states
%  \textit{In situ} measurements of spatial correlations have greatly enhanced the understanding of quantum gases. Spatial correlations can arise from  Bose or Fermi quantum statistics \cite{ngl99,bgm97, kmi97,zhg03,obh15} or formation of more complex entangled states
% %such as in the strongly interacting Tonks-Girardeau regime in a 1D gas
% \cite{bdz08,loh04,kww04}.
 Quantum gas microscopes resolve correlations on length scales on the order of, or larger than
 %equal to or larger than the spacing between sites in an optical lattice, typically half a wavelength of the lattice light,
 a wavelength of  light,
 enabling studies of quantum magnetism \cite{mcj17} and the superfluid-to-Mott insulator transition \cite{bpt10}. Inelastic loss from spin flips and three-body recombination  probe two- and three-body spatial correlations at very short range \cite{bgm97,loh04}.  {Despite the tremendous progress in experimental techniques, \textit{in situ}} probes of spatial correlations between these length scales are {still} lacking. {Many complex many-body phenomena take  place at these intermediate scales, and  such a probe} {would thus provide a new window into}  {physics such as the formation of}  halo dimers \cite{kgj06} and Efimov trimers \cite{cgj10} related to resonant atomic scattering states,  long-range Cooper pairs in strongly interacting Fermi gases \cite{adl04,gps08},  and strongly correlated 1D gases \cite{loh04,kww04}.%, and {generally} the interplay of interactions and quantum statistics % in the pair-correlation function
 %of many-body  bosonic and fermionic systems
 %\cite{ngl99}.

\begin{figure}[t]
\includegraphics[width=0.4\textwidth]{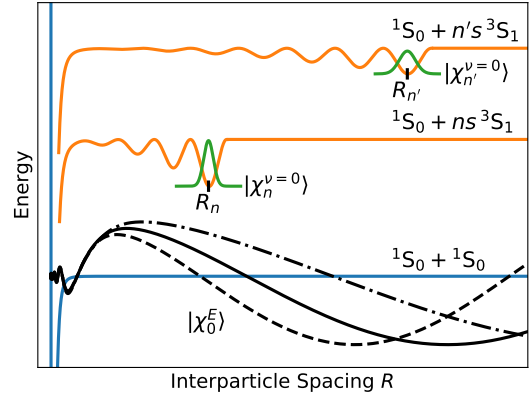}
\caption{ Schematic of the excitation to a Rydberg molecular state $|\chi^{\nu=0}_n\rangle$ {(green)}
in a Rydberg potential {(orange)} from the state of a pair of colliding atoms $|\chi_0^{E}\rangle$ {(black)}. The wavefunction of the ground $\nu=0$ molecular dimer state
 is highly localized in the outer lobe of the molecular potential at $R_n$, as shown for two different principal quantum numbers $n$ and $n^{\prime}$. {Asymptotically far outside the short-range interatomic potential (blue), $|\chi_0^{E}\rangle$  describes  a free particle state with} wave vector $k=\sqrt{2\mu E/\hbar^2}$ for collision energy $E$ and reduced mass $\mu$.   The experiment samples a thermal distribution of collision energies, and the molecular excitation rate is proportional to the pair correlation function $g^{(2)}(R_n)$. %Measuring the excitation rate of $|\chi^{\nu=0}_n\rangle$ at different $n$ provides a probe of $g^{(2)}(R)$ at previously inaccessible length scales.
  }\label{figure:PhotoassociationWavefunctions}
\end{figure}

 Here we demonstrate photo-excitation of ultra-long-range Rydberg-molecule (RM) dimers \cite{gds00,bbn09,dad15} in an ultracold gas as an \textit{in situ} probe of nonlocal pair-correlations \cite{ngl99} at previously inaccessible length scales.
 %In RMs,  one or more ground-state atoms are bound to an atom in a highly excited Rydberg state through scattering between the Rydberg electron and ground-state atoms \cite{gds00,Fermi1934} (Fig.\ \ref{figure:PhotoassociationWavefunctions}).
 %Probing correlations with RMs has many advantages: it is an \textit{in situ}, non-destructive probe that can be tuned over previously inaccessible  length scales.
At short range, we observe bunching in a thermal gas of spinless bosonic
$^{84}$Sr and Pauli exclusion, or anti-bunching, in a polarized gas of
fermionic $^{87}$Sr atoms, reflecting the effects of (anti-)symmetrization of
the wave functions dictated by the spin statistics theorem. These
correlations vanish at distances greater than the thermal de Broglie
wavelength. Bunching and anti-bunching have been observed before in quantum
gases with destructive measurement schemes
\cite{bgm97,shp05,jmh07,ysh96,rbo06,obh15}.{ In contrast, RM excitation
can be nearly non-destructive \cite{mnt15}. It can also probe
 the temporal evolution} of correlations since
the molecular binding energy, and {therefore the
inverse excitation time scale, are much greater than {the relevant many-body energy scales of quantum gases, such as the Fermi energy or} chemical
potential.}

In a RM dimer,  one ground-state atom is bound to  a highly excited Rydberg atom.
The binding potential  results from scattering between the Rydberg electron and ground-state atom \cite{gds00,fer34}, and it {therefore} follows the  Rydberg-electron probability distribution. (See Fig.\ \ref{figure:PhotoassociationWavefunctions}.) For Sr, the atom-electron interaction is attractive, leading to formation of RMs. The molecular potential is also attractive for Rydberg excitations in Rb and Cs quantum gases \cite{btf01,bbn09,bry15}, which attests to the broad applicability of the probe we investigate in the present work.
%In an RM dimer,  one ground-state atom is bound to an atom in a highly excited Rydberg state.
%The binding potential  results from scattering between the Rydberg electron and ground-state atom \cite{gds00,fer34}, and it follows the  Rydberg-electron probability distribution (Fig.\ \ref{figure:PhotoassociationWavefunctions}). For Sr, the atom-electron interaction is attractive, leading to formation of RMs.
%In an RM dimer,  one ground-state atom is bound to an atom in a highly excited Rydberg state through scattering of the Rydberg electron from the ground-state atom \cite{gds00,fer34}. The binding potential follows the  Rydberg-electron probability distribution (Fig.\ \ref{figure:PhotoassociationWavefunctions}). For Sr, the atom-electron interaction is attractive, leading to formation of RMs.

To probe spatial correlations, we exploit the fact that the internuclear separation in the most deeply bound  \hbox{RM} dimer state, $|\chi_n^{\nu=0}\rangle$, is highly localized in the potential minimum formed by the outer lobe of the Rydberg wavefunction located at a separation $R_n\approx 2 (n-\delta)^2 a_0$ (Fig.\ \ref{figure:PhotoassociationWavefunctions}). The quantum defect is $\delta=3.37$ for the $5sns\,^3\text{S}_1$ states used in this work, and $a_0\approx0.05$\,nm is the Bohr radius.
In a simple semi-classical picture, the formation of a molecule
requires the presence of atoms separated by approximately $R_n$. Thus the
excitation rate serves as a measure of the relative probability of finding two
particles with separation $R_n$ in the initial gas,
which can be quantified by the nonlocal pair-correlation function $g^{(2)}{(R_n)}$. For principal quantum number $n$ between 20 and 75, $R_n$ ranges from \hbox{400} to \hbox{$10^4$ $a_0$}, providing an \textit{in situ} probe of correlations at  previously inaccessible length scales. %A probe of correlations in this range provides a new path for exploring spatial correlations in many-body systems.
This method is similar to the mapping of  short-range ($\lesssim 100\,a_0$) atomic scattering states with photoassociative spectroscopy of low-lying energy levels  \cite{nww94,amg96,twj96,bav00,jtl06}. The possibility of measuring non-local correlations with RMs was mentioned in \cite{eil18}, and  short-range correlations were probed with RM excitation in \cite{mnt15}.

In the present work, non-degenerate quantum gases of spin-polarized, fermionic $^{87}$Sr ($I=9/2$) and  bosonic $^{84}$Sr ($I=0$) are used to measure the effects of quantum statistics on the excitation rate of RMs and thus on the pair correlation function.{ As a reference, we employ an unpolarized sample of $^{87}$Sr, which provides a good approximation to a gas of uncorrelated particles because of
 its tenfold-degenerate ground state. } Accordingly, the excitation rates of RMs in $^{84}$Sr and spin-polarized $^{87}$Sr are compared to those of unpolarized $^{87}$Sr to extract $g^{(2)}(R)$.

Ultracold samples of {bosonic} $^{84}$Sr are produced by loading a magneto-optical trap (MOT) operating on the \hbox{$5s^2\,^1\text{S}_0\rightarrow 5s5p\,^1\text{P}_1$} transition at 461 nm. Weak spontaneous decay from the $5s5p\,^1\text{P}_1$ state transfers atoms into the metastable $5s5p\,^3\text{P}_2$ state, a fraction of which are magnetically trapped by the quadrupole magnetic field of the MOT \cite{nsl03}. Magnetically trapped atoms are then returned to the ground state using optical pumping on the $5s5p\,^3\text{P}_2\rightarrow 5p^2\,^3\text{P}_2$ transition at \hbox{481 nm} \cite{hnc18}. Atoms are further cooled to $\sim 2$ $\mu$K using a second MOT operating on the $5s^2\,^1\text{S}_0\rightarrow 5s5p\,^3\text{P}_1$ intercombintation line ($\Gamma/2\pi =7.5$ kHz) at 689 nm. Subsequently, the atoms are  loaded into an optical dipole trap (ODT) formed using 1064 nm light and evaporatively cooled to the desired temperature. %The magnetic field is set to zero for bosonic samples during the ODT and Rydberg excitation stage.

To produce samples of {fermionic} ${^{87}}$Sr,  metastable $5s5p\,^3\text{P}_2$ $^{84}$Sr and $^{87}$Sr atoms are sequentially loaded into the magnetic trap. Both isotopes are repumped simultaneously using 481\,nm light,
%The hyperfine structure and the isotope shift of the repumping transitions are covered by dithering the laser current and using an electro-optic modulator (EOM) to generate spectral sidebands.
after which they are further cooled to $\sim 2$ $\mu$K in a simultaneous dual-isotope narrow-line MOT and loaded into the ODT.

For measurements involving spin-polarized $^{87}$Sr, a bias magnetic field of 7.6 G is applied after loading the ODT, which produces a Zeeman splitting of $\sim650$ kHz between adjacent magnetic sublevels in the \hbox{$5s5p\,^3\text{P}_1\,F=9/2$} manifold. Population is transferred into the $m_F
=9/2$ ground state by applying a series of $\sigma^+$ polarized 689 nm laser pulses approximately 50 kHz red-detuned from each $m_F \rightarrow m_F+1$ transition. Once this optical pumping is complete, the field is lowered to \hbox{$\sim 1$ G} to maintain the quantization axis.  Experiments with unpolarized  samples are performed in zero magnetic field.

After a $^{87}$Sr sample  is prepared in the appropriate state, it is evaporatively cooled  by lowering the optical trap depth over a period of several seconds. The presence of $^{84}$Sr allows for sympathetic cooling of polarized $^{87}$Sr atoms, and yields colder samples of unpolarized $^{87}$Sr atoms. Once the final trap depth is reached, any remaining $^{84}$Sr atoms are removed by scattering light resonant with the $5s^2\,^1\text{S}_0\rightarrow\,5s5p\,^3\text{P}_1$ transition. The isotope shift between $^{87}$Sr and $^{84}$Sr {ensures} that no significant heating of the $^{87}$Sr atoms occurs.

\begin{figure}[t]
\includegraphics[width=0.4\textwidth]{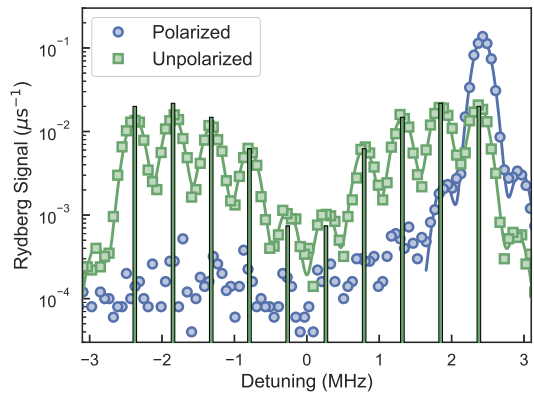}
\caption{Spectra for excitation to the atomic $n=34$ Rydberg state for spin-polarized (circles) and unpolarized (squares) gases of $^{87}$Sr. A 1 G magnetic field causes the observed Zeeman splitting. The vertical bars indicate the square of the product of Clebsch-Gordan coefficients associated with each transition, and differences with the measured peak heights point to small
%The relative heights of the Zeeman peaks are given by the Clebsch-Gordan coefficients (bars). Discrepancies between peak heights and the size of the Clebsch-Gordan coefficients indicate
deviations from an equal distribution of $m_F$ levels in the ground state. Curves show fits used to extract the population in each $m_F$ level. The small features on the extreme right and extreme left of the plot arise due to imperfect polarization of the first photon, and are included in the model for completeness.\label{figure:polarizationClebsch}}
\end{figure}

\begin{figure*}[t]
\includegraphics[width=0.9\textwidth]{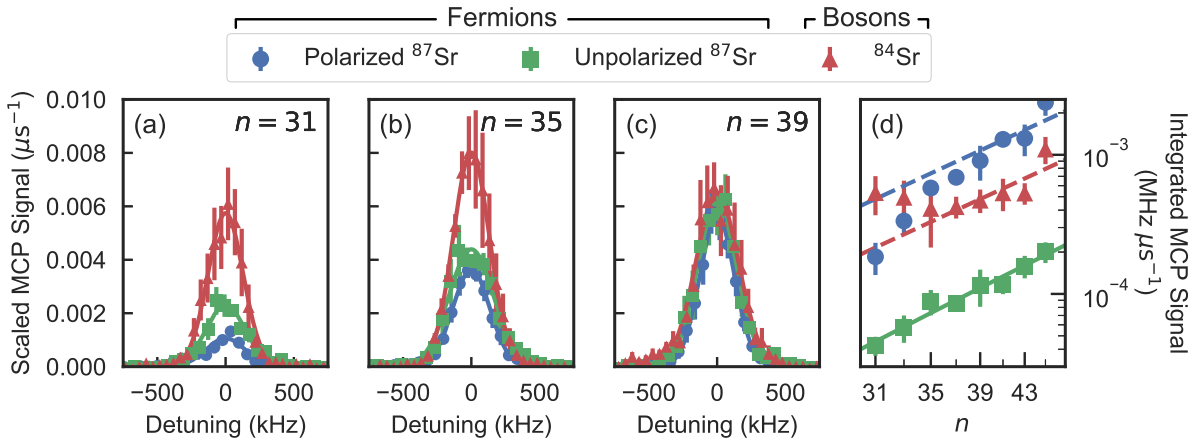}%{polarized_vs_unpolarized_linear_scale_with_fits.eps}
\caption{ Raw data showing the effects of quantum statistics on the
  excitation of RMs. (a-c) Spectra for excitation to the
  $|\chi_n^{\nu=0}\rangle$ dimer ground state for a spin-polarized $^{87}$Sr
  Fermi gas (blue, circles), for an unpolarized $^{87}$Sr Fermi gas (green,
  squares) and a spinless $^{84}$Sr Bose gas (red, triangles). The spectra for
  the polarized Fermi gas and the spinless Bose gas are scaled such that the
  spectra match at $n=39$ to highlight the effects of quantum statistics (see
  text). (d)  {Integrated signal of the RM dimer} spectra versus principal quantum number. A fit
  (solid green line)  shows that the integral for the unpolarized gas varies as
  {$(n-\delta)^{\alpha}$, with $\alpha = 3.5(3)$}, reflecting variation in the $n$-dependent Franck-Condon factors and electronic matrix element, with some contribution from faster natural decay at lower $n$. The dashed lines are translations of the $(n-\delta)^{\alpha}$ curve to guide the eye.
% to highlight
%deviations from this power-law scaling for polarized fermions and bosons, which reflect
% the effect of quantum statistics.
	  }\label{figure:RawData}
\end{figure*}

A two-photon transition is employed to create Rydberg atoms or molecules. The first photon, at 689 nm, has a fixed blue-detuning of $14$ MHz from the $5s5p\,^3\text{P}_1$ level ($F=11/2$ for $^{87}$Sr). The energy of the second photon, at 320 nm,  is scanned to obtain spectra for excitation to $5sns\,^3\text{S}_1$ levels ($F=11/2$ for $^{87}$Sr) (Figs.~\ref{figure:polarizationClebsch} and \ref{figure:RawData}). The excitation lasers are applied for 10 $\mu$s, after which an electric field is applied to ionize   Rydberg excitations. Product electrons  are counted using a microchannel plate  detector. Typically, 1000 laser pulses {at a single frequency are applied to each sample}.

To quantitatively measure the pair-correlation function $g^{(2)}(R)$, {the molecular excitation rate in a correlated gas is normalized with respect to the rate in an uncorrelated gas. This allows us to cancel experimental factors and, most importantly, $n$-dependent contributions to the excitation rate that are unrelated to spatial correlations. Maintaining similar sample densities and temperatures  increases the accuracy of this procedure. To this end, we approximately match the final trap potential, excitation laser intensity, atom number, peak density, and sample temperature. The latter quantities are {inferred from} time-of-flight absorption imaging on the $5s^2\,^1\text{S}_0\rightarrow\,5s5p\,^1\text{P}_1$ transition at 461\,nm and knowledge of the trapping potential \cite{ssk14}.

The effectiveness of optical pumping is measured  spectroscopically through
excitation of the $5sns\,^3\text{S}_1$ $F=11/2$ atomic Rydberg state in a 1 G
magnetic field using two $\pi$-polarized photons ($\Delta m_F=0$). The
spectrum  (Fig.\ \ref{figure:polarizationClebsch}) is fit to an appropriate
lineshape model to determine the degree of polarization. For a polarized sample, at least 90\% of the atoms occupy the $m_F=9/2$ state. For the unpolarized Fermi gas,  the populations in the ten different ground-state $m_F$ levels are approximately equal ($\pm 25\%$). The differing heights of the Zeeman peaks in the  spectrum for the
unpolarized sample arise from the variation of transition
strength from the ground state to final states with different magnetic quantum
numbers as given by angular momentum coupling (Clebsch-Gordan coefficients).  %A fit to the  spectrum for a polarized sample shows that at least 90\% of the atoms occupy the $m_F=9/2$ state.

To probe spatial correlations and measure $g^{(2)}(R)$, a polarization configuration different from the one used to probe polarization is used to create Rydberg molecules. The first photon is $\sigma^+$ polarized, and the second photon remains $\pi$ polarized, which maximizes the transition strength for the spin-polarized sample.
%The detuning of the 689 nm photon is large compared to the natural linewidth of the transition, but some off-resonant scattering can still occur. However, because this photon has $\sigma^+$ polarization, scattering does not decrease the ground-state spin polarization.

\begin{comment}
In the fermion (anti-bunching) experiment we load $2.2\times 10^5$ $^{87}$Sr atoms at 810 nK and 860 nK for the unpolarized and polarized samples respectively. The Fermi temperature $T_{\TEXT{F}}$ is sufficiently low for each sample, $T/T_{\TEXT{F}}=0.98$ for the polarized gas and $T/T_{\TEXT{F}}=1.9$ for the unpolarized gas, that we treat the spatial distribution of the gas with Maxwell-Boltzmann statistics and calculate the peak density to be $2.3\times 10^{13}\,\text{cm}^{-3}$ for the unpolarized gas and $2.2\times 10^{13}\,\text{cm}^{-3}$ for the polarized gas.

In the boson (bunching) experiment we load $2.1\times 10^5$ $^{84}$Sr and $2.2\times 10^5$ $^{87}$Sr atoms. The temperatures are slightly lower than the fermion experiment, 650 nK for the Bose gas and 680 nK for the unpolarized Fermi gas. The Bose gas is well above the condensation temperature, $T/T_{\text{c}}=1.4$. The unpolarized Fermi gas is also non-degenerate, $T/T_{\TEXT{F}}=1.7$. Since both gases are sufficiently far from degeneracy we treat the samples using Maxwell-Boltzmann statistics. The peak density of the Bose gas is $3.0\times 10^{13}\,\text{cm}^{-3}$, the peak density of the unpolarized Fermi gas is $3.1\times 10^{13}\,\text{cm}^{-3}$.
\end{comment}

The influence of quantum statistics on spatial correlations is readily apparent in the spectra for excitation to the $\nu=0$ RM state at principal quantum numbers $31\leq n \leq 45$ \hbox{($1400 \,a_0\leq R_n \leq 3200 \,a_0$)}. Figures \ref{figure:RawData}(a-c) show spectra for $n=31,35,39$. The spectra for the spin-polarized Fermi gas ($T=650$\,nK) and Bose gas ($T=860$\,nK) are scaled such that %by 0.22 and 0.11 respectively to
their integrals match the integral of the unpolarized data ($T=860$\,nK) at
$n=39$, where the effects of quantum statistics are small
[Fig.\ \ref{figure:RawData}(c)].  For decreasing quantum number [Fig.\ \ref{figure:RawData}(a,b)], the suppression of the excitation rate in the spin-polarized Fermi gas arising from Pauli exclusion and the enhancement for the Bose gas due to bunching are evident.

%%%%%%%%%%%%%%%%%%%%%%%%%%%%%%%%%%%%%%%%%%%

Figure \ref{figure:RawData}(d) shows the integral of the molecular signals measured for each principal quantum number. The integrals for the unpolarized Fermi gas (green circles) can be fit well by an $(n-\delta)^{3.5}$ power law. In the absence of effects of quantum statistics, the integrals for all samples should have the same $n$ dependence but different overall amplitudes that
reflect $n$-independent factors such as Clebsch-Gordan coefficients (Fig.\ \ref{figure:polarizationClebsch}), differences in detector efficiency arising from the magnetic field needed to preserve quantization for the spin-polarized sample, and differences in laser intensity between the Bose and spin-polarized Fermi gas experiments.
Deviations from the $(n-\delta)^{3.5}$ power law at low quantum number result from quantum statistics.
%, and slightly different temperatures and densities between samples.
 %The dashed lines follow the $n^{3.8}$ dependence of the fitted line, but are scaled to highlight deviations at low quantum number due to the effects of quantum statistics.
 %The overall scale factor originates from Clebsch-Gordan coefficients, differences in detector efficiency, differences in laser intensity between the Bose and spin-polarized Fermi gas experiments, and slightly different temperatures and densities between samples. However, these effects are constant with $n$ and do not affect the $n$ dependent physics.

{{
The excitation probability to the ground vibrational state ($\nu=0$) of the RM for principal quantum number $n$  is proportional to a Franck-Condon factor  that accounts for  a thermal average over collision energy ($\langle ...\rangle_E$) for initial two-particle states and all possible initial and final rotational states. This reduces to
$ {\cal F}_n= \sum_{l}(2l+1)\langle |\int dR R^2 \chi_n^{\nu=0}(R) \chi_0^{E,l}(R)|^2 \rangle_E$  \cite{wag19}, where $\chi_n^{\nu=0}$ is the radial wavefunction for the RM, which is independent of $l$ for the low-$l$ states  contributing to ${\cal F}_n$.  $\chi_0^{E,l}$ is the wavefunction for the initial state with collisional energy $E$ and rotational angular momentum quantum number $l$.  The sum over $l$ is understood to be restricted to initial states with allowed exchange symmetry.

$\chi_n^{\nu=0}$ is well-localized at $R_n$ on the scale of the initial collisional state.
In particular,  the wave
function  for $n < 50$  is localized within a single potential well
(Fig.\ \ref{figure:PhotoassociationWavefunctions}).
This allows ${\cal F}_n $
%for colliding particles initially in internal spin states $i$ and $j$
to be approximated as \cite{wag19}
%\begin{equation}
%{\cal F}_n
%\simeq g_{i,j}^{(2)}(R_n)
%  \left| \int dR R^2 \chi_n^{\nu=0}(R)  \right|^2
%\equiv g_{i,j}^{(2)}(R_n) \, \mathcal{O}_n
%\label{eq:Franck-Condon}
%\end{equation}
\begin{equation}
{\cal F}_n
\simeq 
  \left| \int dR R^2 \chi_n^{\nu=0}(R)  \right|^2 g^{(2)}(R_n)
\equiv   \mathcal{O}_n \, g^{(2)}(R_n)
\label{eq:Franck-Condon}
\end{equation}
where $g^{(2)}(R_n)$ is the pair correlation function for separation $R_n$, and $\mathcal{O}_{n}$ is an effective Franck-Condon factor.
This derivation can be generalized to the case of an initial state of a many-body Fermi or Bose gas at arbitrary density and temperature  and with multiple internal spin states initially populated.
}
}

{{
When experimental factors are taken into account,
%one can show \cite{wag19} that  for excitation
%of indistinguishable particles initially in internal states $i$ and $j$
%to the ground-state dimer with principal quantum number $n$,
the integrated  signal %for polarized fermions/bosons or unpolarized fermions
becomes
% \begin{equation}
%\mathcal{S}_{n}
%\simeq
%%\alpha \int\,d^3r\,df \sum_{r,b} \Gamma^{\gamma}_{\nu,r,b}
% \alpha  I_1 I_2 \mathcal{N} \beta_{n} \mathcal{O}_{n}\times \left\{
%\begin{array}{cc}
%  C_{\mathrm{pol}}\,  g^{(2)}_{i,i}(R_n) & \mathrm{polarized} \\
%   C_{\mathrm{unpol}} \,  g^{(2)}_{i\neq j}(R_n) & \mathrm{unpolarized}
%\end{array}
%  \right.
%\label{eq:Signal}
%\end{equation}
 \begin{equation}
\mathcal{S}_{n}
\simeq
%\alpha \int\,d^3r\,df \sum_{r,b} \Gamma^{\gamma}_{\nu,r,b}
 \alpha  I_1 I_2 \mathcal{N} \beta_{n} 
  \mathcal{O}_{n} C g^{(2)}(R_n),
\label{eq:Signal}
\end{equation}
which is proportional  to
the detector efficiency $\alpha$, the two-photon-excitation laser intensities $I_1$ and $I_2$, the volume integral of the square of the density distribution $\mathcal{N}\equiv\int d^3r \rho(\mathbf{r})^2$, a factor $\beta_{n}$ proportional to the square of the reduced two-photon electronic-transition matrix element,  a factor $C$ expressible in terms of Clebsch-Gordan coefficients, $\mathcal{O}_{n}$, and $g^{(2)}(R_n)$.
}}

For non-degenerate gases of noninteracting particles, $g^{(2)}(R)$ should be given by
\begin{equation}\label{eq:thermalpaircorrelationfunction}
 g^{(2)}(R)=1+\epsilon\, \mathrm{e}^{-2\pi R^2/\lambda_{\text{dB}}^2},
\end{equation}
where $\lambda_{\text{dB}}=h/\sqrt{2\pi mk_BT}$ is the thermal de Broglie wavelength, and $\epsilon$ equals $+/-$\,1 for indistinguishable thermal bosons/fermions in identical internal states
and 0 for  classical statistics \cite{ngl99,dlm18}. Trap and phase-space-density-dependent corrections to Eq.\ (\ref{eq:thermalpaircorrelationfunction}) will vary with separation $R$ and are always less than $z/10$  \cite{ngl99}, where  $z\approx \rho\lambda_{\text{dB}}^3$ is the fugacity. The highest peak fugacity of any of the samples used in these experiments is $z=0.4$, and corrections are small. Eq. (\ref{eq:thermalpaircorrelationfunction}) neglects interactions between ground-state particles, which modify spatial correlations at length scales less than the scattering length or the range of the ground-state atom-atom molecular potential, which are much smaller than $R_n$ probed in this experiment.

Equation~(\ref{eq:Signal}) is used to experimentally determine $g^{(2)}(R_n)$ for indistinguishable particles in identical internal states by normalizing the integrated signals for the bosons and spin-polarized fermions to the integrated signal for the unpolarized fermions. This cancels common factors  $\beta_{n}$ and $\mathcal{O}_{n}$. For the unpolarized Fermi gas, we assume $g^{(2)}(R)=1-0.1\mathrm{e}^{-2\pi R^2/\lambda_{\text{dB}}^2}$, which is the generalization of Eq. (\ref{eq:thermalpaircorrelationfunction}) for equal populations in the ten ground spin-states.
The remaining factors that vary between different experimental runs and different isotopes and sample polarizations, are either measured or calculated independently.
 The Clebsch-Gordan factor for the unpolarized gas ($C_{\mathrm{unpol}}$) is calculated assuming equal populations in all ground spin states, yielding
$C_{\mathrm{pol}}/C_{\mathrm{unpol}}=5.05$, where $C_{\mathrm{pol}}$ describes bosons and polarized fermions.
Temperatures and densities (peak density $\sim 3\times 10^{13}$\,cm$^{-3}$) of each sample and the unpolarized gas used for normalization match within 10\% in all cases. For the Bose gas, $T/T_{\text{c}}\approx1.5$ where $T_{\text{c}}$ is the critical temperature for Bose condensation. For the polarized Fermi gases, $T/T_{\text{F}}\approx1.0$, where $T_{\text{F}}$ is the Fermi temperature. For the unpolarized Fermi gases, $T/T_{\text{F}}\approx 2$. Density distributions are calculated using the appropriate Bose or Fermi distributions.

At large separations, where the effect of quantum statistics should be negligible, this procedure yields $g^{(2)}(R)=1.5$ rather than the expected value of 1. A systematic deviation of this size is consistent with uncertainties in relative populations of the initial internal spin states of the unpolarized Fermi gas and in trap geometry and resulting density profiles. Ratios are thus divided by an additional correction factor of 1.5 to obtain the values of $g^{(2)}(R_n)$ in Fig.\ \ref{figure:g2}.

  \begin{figure}[t]
\includegraphics[width=0.4\textwidth]{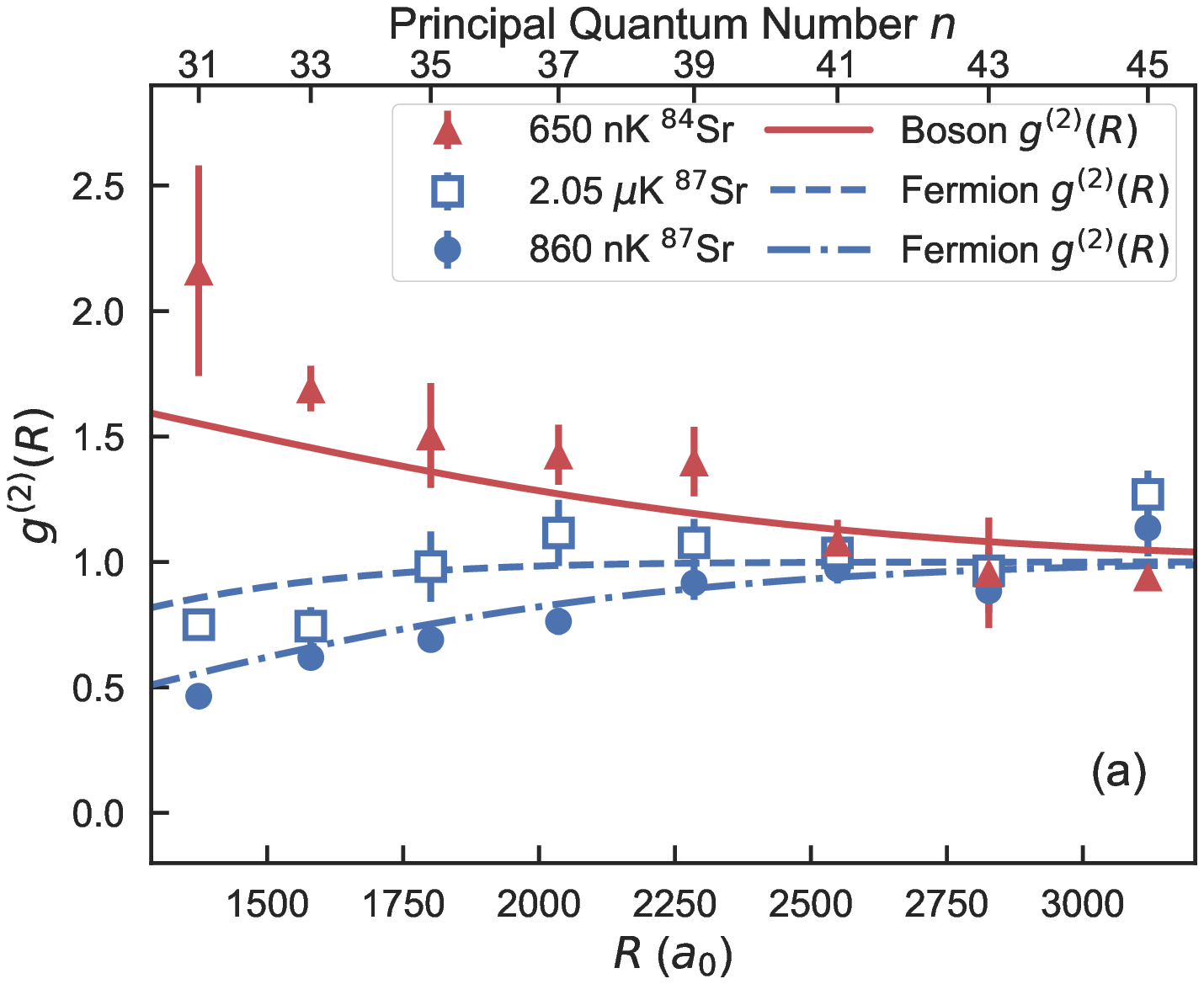}
\includegraphics[width=0.4\textwidth]{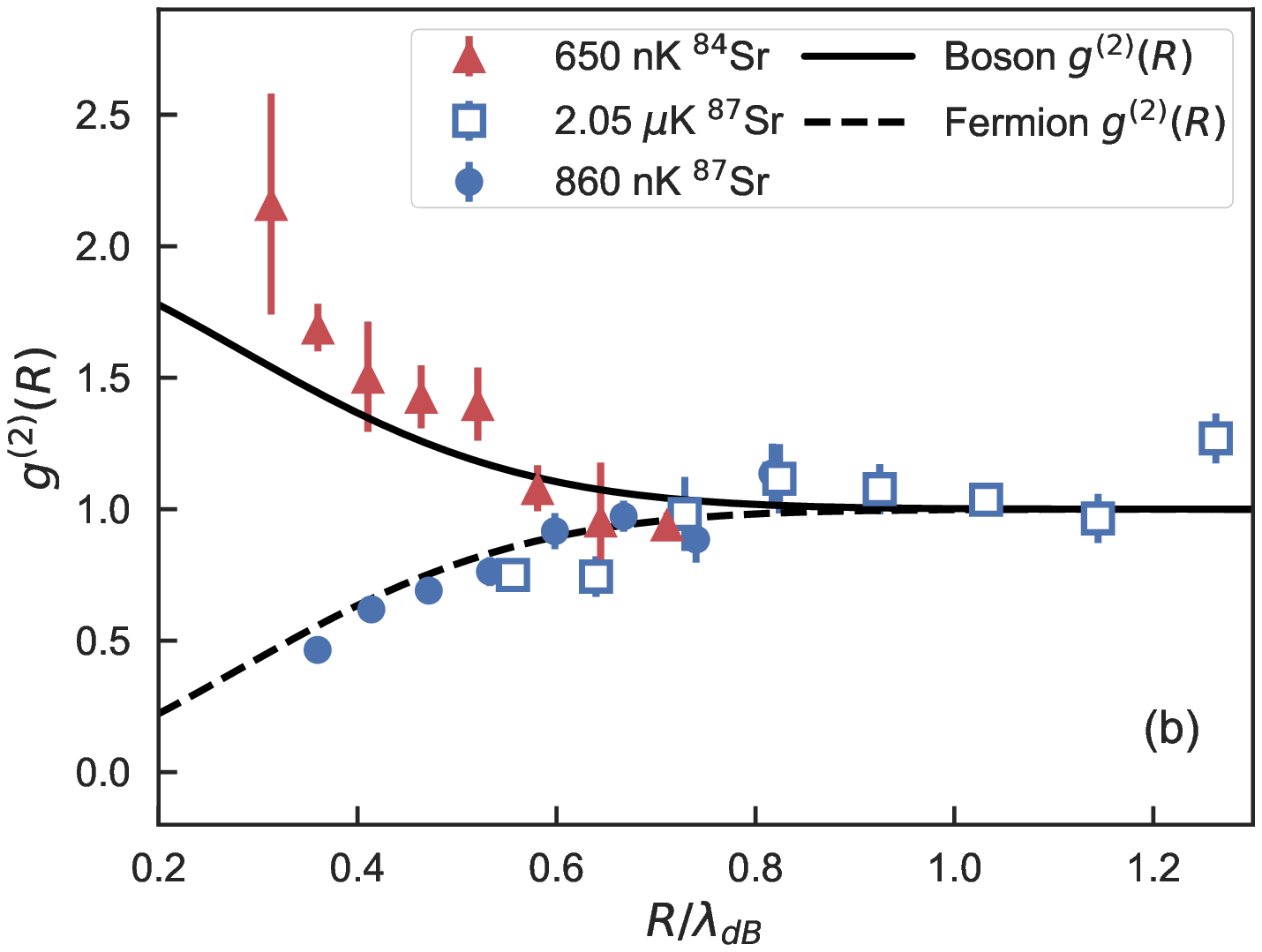}
\caption{Measured pair-correlation function for indistinguishable particles in identical internal states.   (a) $g^{(2)}(R)$ for a Bose gas and  Fermi gases at two different temperatures plotted against interparticle separation, $R_n$. Sample temperatures are indicated in the legend. (b) $g^{(2)}(R)$ for Fermi and Bose gases plotted against $R_n$ scaled by the thermal de Broglie wavelength. The two sets of fermion measurements (blue symbols) fall onto a single curve and approach a constant value at large scaled distances. Error bars indicate statistical fluctuations from repeated measurements. Expected $g^{(2)}(R)$ (Eq.\ (\ref{eq:thermalpaircorrelationfunction})) for bosons (solid) and fermions (dashed and dot-dashed) are shown by the lines.
  }\label{figure:g2}
\end{figure}

At lower values of $n$ the normalized integrated signals for the bosons and fermions clearly deviate from unity  (Fig.\ \ref{figure:g2}(a)). The boson signal increases while the fermion signal decreases, which is consistent with bunching and anti-bunching respectively.
The Fermi-gas experiment was performed at two different temperatures, and anti-bunching is less pronounced in the warmer sample reflecting the shorter thermal de Broglie wavelength.
%
%
  %A first-principle description of RM dimer excitation, including the effects of quantum statistics, is currently lacking.
%  However, neglecting small corrections from anti-bunching for collisions between atoms in the same spin state in the unpolarized Fermi gas used for normalization, we expect  the signal ratios described above to be proportional to $g^{(2)}_{i,i}(R_n)$, the pair correlation function between indistinguishable bosons or fermions in identical internal states at a separation determined by the radius of the outer lobe of the Rydberg orbital, $R_n$. At large separations, where the effect of quantum statistics should be negligible, we find that the ratios approach $1.4$ rather than the expected value of 1. A systematic error of this size is consistent with uncertainties in relative populations of the ground states of the unpolarized Fermi gas and in trap geometry and resulting density profiles. The ratios are corrected this factor to obtain the values of $g^{(2)}_{i,i}(R_n)$ in Fig.\ \ref{figure:g2}.
%
%Figure \ref{figure:g2}(a) clearly shows bunching for bosons and anti-bunching for polarized fermions.
%The Fermi gas experiment was performed at two different temperatures, and the anti-bunching is less pronounced in the warmer sample reflecting the shorter thermal de Broglie wavelength, $\lambda_{\text{dB}}=h/\sqrt{2\pi mk_BT}$.
%
Figure \ref{figure:g2} also shows that the results  follow the expected behavior for   $g^{(2)}(R)$ (Eq.\ (\ref{eq:thermalpaircorrelationfunction})).
On a scaled, dimensionless axis (Fig.\ \ref{figure:g2}(b)), the two fermionic data sets fall on the same curve and approach a constant value for larger $R/\lambda_{\text{dB}}$. Effects of quantum statistics - bunching for bosons and anti-bunching due to Pauli exclusion for fermions - are strikingly evident.
%, justifying the introduction of $g_{ij}^{(2)}$ in Eq.\ \ref{eq:Signal}.

In summary, we have demonstrated that photoexcitation of the most deeply bound, $\nu=0$ dimer RM state provides an \textit{in situ} probe of pair-correlations in an ultracold gas that can be tuned over previously inaccessible  length scales.
%There are many interesting prospects to consider for this diagnostic.
These results suggest other interesting phenomena that can be studied with this diagnostic.
%that RMs can be used to probe the effects of  interactions on the interparticle correlation function.
For example, the pair-correlation function in a  gas with a large s-wave scattering length, in the range of $R_n$ probed in this experiment, should show strong deviations from the non-interacting result presented in Eq. (\ref{eq:thermalpaircorrelationfunction}).
%The effects of quantum statistics should also be visible in the excitation rate of vibrationally excited dimer states, which are much less localized but probe smaller interparticle separations.
Stronger suppression/enhancement effects on higher-order correlations should be observable with trimers, tetramers, etc.
Moreover, due to the fact that Rydberg molecule formation takes place on a time scale ($\sim 1\,\mu\mathrm{s}$) much faster than the relevant  many-body dynamics of quantum gases, RMs hold promise for \textit{in situ} probing of the time evolution of correlations during the non-equilibrium dynamics following quantum quenches or in driven many-body systems.

%In principle, RM excitation can be nearly non-destructive \cite{mnt15}.
%A first-principle derivation of the dimer RM excitation rate, including the effects of quantum statistics, would be of great value to better understand the limitations of this new probe.

%%%%%%%%%%%%%%%%%%%%%%%%%%%%%%%%%%%%%%%%%%%%%%%%%%

\textbf{Acknowledgements}:
Research supported by the AFOSR (FA9550-14-1-0007), the NSF (1301773, 1600059, and 1205946), the Robert A, Welch Foundation (C-0734 and C-1844), the FWF(Austria) (FWF-SFB041 ViCoM, and FWF-SFB049 NextLite).  The Vienna scientific cluster was used for the calculations.  H.~R.~S. was supported by  the NSF through a grant for the Institute for Theoretical Atomic, Molecular, and Optical Physics at Harvard University and the Smithsonian Astrophysical Observatory. R. S. and M. W. were supported by the Deutsche Forschungsgemeinschaft (DFG, German Research Foundation) under Germany’s Excellence Strategy – EXC-2111 – 390814868. %All plots were produced using the matplotlib graphics environment \cite{Hunter:2007}.

%Control: production of eprint (-1) disabled
%

%\bibliography{bibliography,library}
%\bibliography{../../PAPERS/BIBLIOGRAPHY/bibliography}
%\bibliography{../../PAPERS/BIBLIOGRAPHY/bibliography,../../PAPERS/BIBLIOGRAPHY/MENDELAYTEMPLIBRARYBIBFILE/library}

\end{document}